\documentclass[epj,referee]{svjour}
\usepackage{graphics}
 
\usepackage{amssymb}
\usepackage{verbatim}
\usepackage[square,comma]{natbib}

\newcommand{\bea}{\begin{eqnarray}}
\newcommand{\be}{\begin{equation}}
\newcommand{\ee}{\end{equation}}
\newcommand{\eea}{\end{eqnarray}}
\newcommand{\bfi}{\begin{figure}}
\newcommand{\efi}{\end{figure}}
\newcommand{\bc}{\begin{center}}
\newcommand{\ec}{\end{center}}
\newcommand{\nn}{\nonumber}

\begin{document}
\title{Roughness and Finite Size Effect in the NYSE
Stock-Price Fluctuations}
\author{V. Alfi\inst{1,2} \and F. Coccetti\inst{3} \and A. Petri \inst{4}
\and L. Pietronero \inst{1,4}}
\institute{``La Sapienza'' University, Physics Department, 
  P.le A. Moro 5, 00185, Rome, Italy \and ``Roma Tre'' University, Physics
  Department,  V. della Vasca Navale 84, 00146, Rome, Italy \and 
Museo Storico della Fisica e Centro Studi e Ricerche ``Enrico Fermi``, Via
Panisperna, Roma, Italy \and Istituto dei
  Sistemi Complessi - CNR, Via Fosso del Cavaliere 100, 00133 Roma, Italy }
\date{Received: 30 January 2006}
\abstract{
We consider the roughness properties of NYSE (New York Stock Exchange) stock-price fluctuations. The
statistical properties of the data are relatively homogeneous within the same
day but the large jumps between different days prevent the extension of the
analysis to large times. This leads to intrinsic finite size effects which
alter the apparent Hurst (H) exponent.
We show, by analytical methods, that finite size effects always lead to an
enhancement of H. We then consider the effect of fat tails on the analysis
of the roughness and show that the finite size effects are strongly enhanced
by the fat tails. The non stationarity of the stock price dynamics also
enhances the finite size effects which, in principle, can become important even in the
asymptotic regime. We then compute the Hurst exponent for a set of stocks of
the NYSE and argue that the interpretation of the value 
of H is highly ambiguous in view of the above results. Finally we propose an
alternative determination of the roughness in terms of the fluctuations from
moving averages with variable characteristic times. This permits to eliminate
most of the previous problems and to characterize the roughness in useful
way. In particular this approach corresponds to the automatic elimination of
trends at any scale.
}
\maketitle

\section{Introduction}

The dynamics and fluctuations of stock-prices is
represented, at the simplest level, by a random walk which
guarantees for the basic property of an efficient market.
In the past years it has become clear that one faces a rather
subtle and complex form of random walk.
Simple correlations of price change are indeed zero at the shortest time 
but many other features, often related to power law behavior have been 
discovered~\citep{bib1}.
Among the most preeminent one may mention the power law distributions of
returns (``fat tails'') and the volatility clustering~\citep{bib2,bib3}.
These properties, however, are far from exhaustive and other
approaches have been introduced in the attempt of describing the
subtle correlations of stock-price dynamics.

One of these methods is the attempt to characterize the ``roughness''
of the dynamics which can provide additional information with respect to the
fat tails and volatility. The scaling properties of the roughness can be
defined via the so called Hurst exponent $H$~\citep{bib3}. 
We consider the roughness problem for
high frequency NYSE stock-prices.
This means that we take into account all the transactions which occur (tick by
tick).

First we discuss the statistical properties of the data set and show
that finite size effects are unavoidable
and very important. Then we show that fat tails and correlations affect the 
value of the Hurst exponent in an important
way.
Finally we analyze the real stock-price fluctuations and argue that the Hurst
exponent alone cannot properly characterize their roughness.
To this purpose we use a new method to study the roughness which
is able to automatically  eliminate the trend problem.
This is based on the deviation from a suitable moving average and it resolves
various  ambiguities of the Hurst approach.

The paper is organized in the following way:

In Section 2 we discuss the database for the high frequency samples.
This will lead to a crucial role of finite size effects because the data are
relatively homogeneous within the same day 
but there is a large gap in price between the closing of one day and the
opening of the next day.

In Section 3 we discuss the general problem associated to the 
determination of the roughness via the Hurst exponent in view of the
anisotropic scaling.

In Section 4 we consider the finite size effects on the roughness exponent 
in random walks with an analytical approach and then include also the possible
effects of fat tails and correlations with Monte Carlo simulations.

In Section 5 we present the roughness analysis for s selection of 
NYSE stock-prices also as a function of time.

In Section 6 we critically analyze the scaling assumption in
relation to the roughness and consider new tools to this
purpose which eliminate the trends at all scales automatically.

In Section 7 we discuss the results and present the conclusions.

\section{Database properties}
We consider as database the price time series of all the transactions of a 
selection of 20 NYSE stocks.
These have been selected to be representative and with intermediate
volatility.
This corresponds to  volumes of $10^5-10^6$
stocks exchanged per day.
We consider 80 days from October 2004 to February 2005.
\begin{figure}[h]
\begin{center}
\resizebox{0.75\columnwidth}{!}{\includegraphics{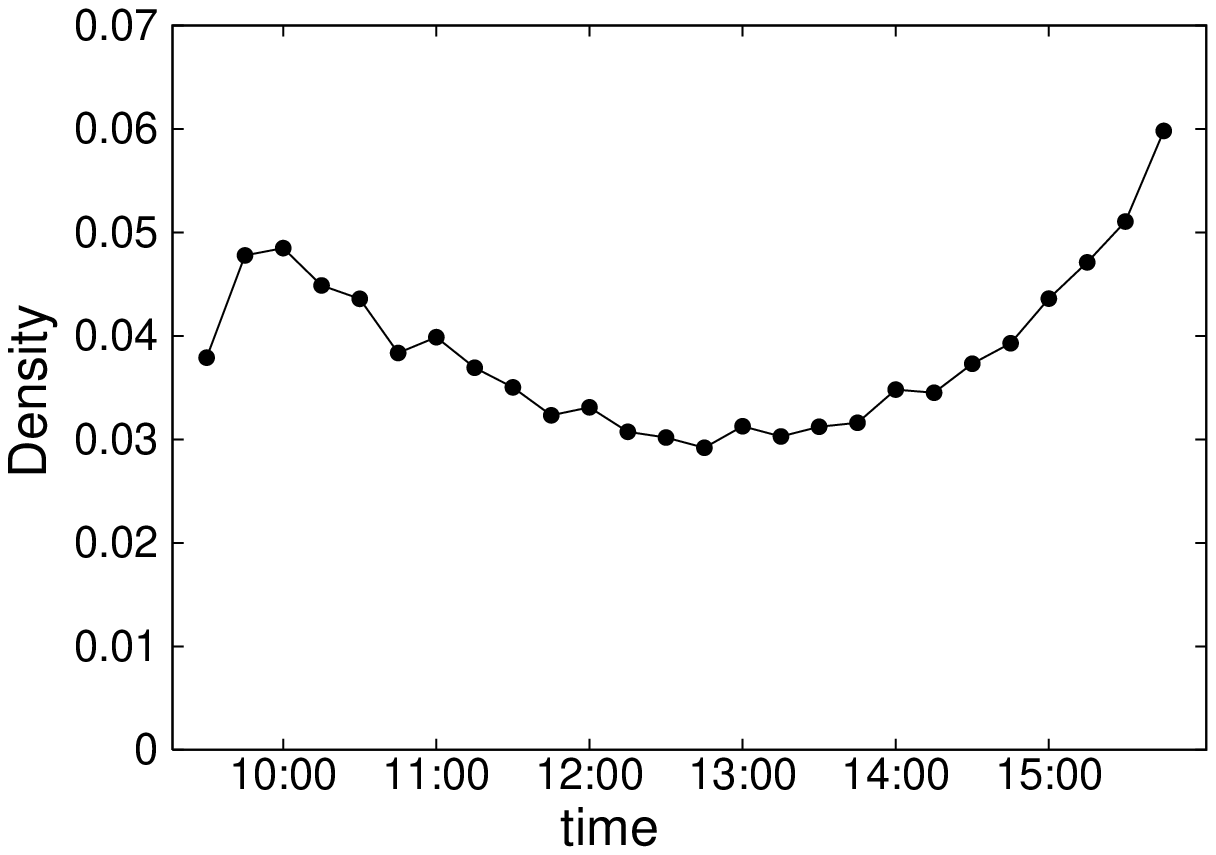}} 
\caption{Behavior of the density of transactions within a day. This concave
  behavior with a maximum fluctuation up to a factor of two is a general
  feature for  all stocks.}
\label{fig1}
\end{center}
\end{figure}

The time series we consider are by a sequential order tick by tick.
This is not identical to the price value as a function of
physical time but we have tested that the results
are rather insensitive to this choice.

The number of transactions per day ranges from 500 to 5000 implying
a typical time interval between transactions of a few seconds.
The density of operations within a day is characterized by a 
concave shape which is rather universal as shown in Fig. \ref{fig1}.
This means that, with respect to the physical time there are 
systematic density fluctuations up to a factor of two with a 
minimum around the center.
This effect is obviously eliminated in our tick by tick
time, in which physical time is not considered and we have tested that it is  not relevant
for the roughness properties.

\begin{figure}[h]
\begin{center}
\resizebox{0.75\columnwidth}{!}{\includegraphics{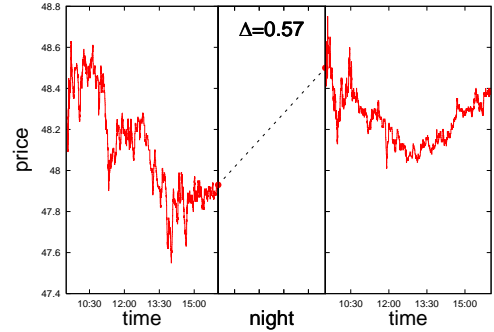}} 
\caption{Night jumps between two days. This gap is very large, typically
of the order of the total daily fluctuation (see Tab.\ref{tab1}).
This poses a serious problem in the analysis of roughness because data
are reasonably homogeneous only within a single day. This leads to the
importance of finite size effects in the analysis.}
\label{fig2}
\end{center}
\end{figure}

A problem which 
is very important and rarely discussed in the literature, 
is the fact that the closing price of a given day is usually
very different than the opening price of the next day.
A typical behavior is illustrated in Fig. \ref{fig2}.
and it shows that these jumps are serious problem in linking the data
of one day to those of the next day.

This means that the data are reasonably homogeneous from
the time scale of a few seconds to a few hours but
going to longer times can be rather arbitrary due to these
large night jumps.

\begin{table}[h]
\caption{Properties of the night jumps with respect to the daily fluctuations
of various stocks. The data refer to the average values over 80 trading days.
$<|p_0-p_c|>$: average absolute value of the gap between opening and closing
price for each day.
$<|\Delta|>$: average of the absolute value of the night jumps.
$\sigma_{\Delta}$: variance of the absolute value of the night jumps.
$\sigma_p$: total daily variance of the price value.
$\sigma_{\delta p}$: variance of the price fluctuations between two
transactions.
}
\label{tab1}
\begin{scriptsize}
\begin{tabular}{lccccc}
\hline\noalign{\smallskip}
stock & $<|p_{o}-p_{c}|>$ &  $<|\Delta|>$  & $\sigma_{\Delta}$ & $\sigma_{p}$  &  $\sigma_{\delta p}$\\ 
\noalign{\smallskip}\hline\noalign{\smallskip}
AH &  0.73494 & 0.36950 & 0.59100 & 0.28539 & 0.02152 \\ 
AVO &  0.47561 & 0.18862 & 0.56508 & 0.23161 & 0.01698 \\ 
BA &  0.41926 & 0.21437 & 0.42131 & 0.19530 & 0.01056 \\ 
BRO &  0.40877 & 0.15750 & 0.37375 & 0.19091 & 0.01607 \\ 
CAI &  0.81284 & 0.39238 & 0.86836 & 0.31750 & 0.02323 \\ 
DRI &  0.30753 & 0.09850 & 0.23245 & 0.11490 & 0.01065 \\ 
GE &  0.22691 & 0.11688 & 0.17154 & 0.10304 & 0.00652 \\ 
GLK &  0.28272 & 0.10212 & 0.23420 & 0.12998 & 0.01054 \\ 
GM &  0.35593 & 0.15725 & 0.25058 & 0.14597 & 0.00833 \\ 
JWN &  0.44531 & 0.23325 & 0.45625 & 0.20444 & 0.01249 \\ 
KSS &  0.57759 & 0.29628 & 0.48844 & 0.22275 & 0.01355 \\ 
MCD &  0.24457 & 0.13850 & 0.20268 & 0.10288 & 0.00758 \\ 
MHS &  0.43605 & 0.20437 & 0.40161 & 0.17267 & 0.01126 \\ 
MIK &  0.34531 & 0.62375 & 3.12751 & 0.14479 & 0.01320 \\ 
MLS &  0.55309 & 0.17287 & 0.27860 & 0.21948 & 0.02045 \\ 
PG &  0.40321 & 0.24462 & 0.46493 & 0.17056 & 0.00906 \\ 
TXI &  0.79704 & 0.22362 & 0.62309 & 0.33799 & 0.02964 \\ 
UDI &  0.44679 & 0.22375 & 0.80100 & 0.19003 & 0.01469 \\ 
VNO &  0.65864 & 0.21950 & 0.36921 & 0.26285 & 0.02443 \\ 
WGR &  0.40877 & 0.16937 & 0.36687 & 0.17846 & 0.01681 \\ 
\noalign{\smallskip}\hline
\end{tabular}
\end{scriptsize}
\end{table}
In Tab. \ref {tab1} we present a detailed analysis of this phenomenon.
For each stock we have in the first column 
the average over 80 days of the absolute value 
of the gap between opening and closing price, $(<|p_o-p_c|>)$,
indicated in US \$.
In the second column we indicate with $<|\Delta|>$ the average 
of the absolute values of the night jumps.
One can see immediately that they are of the same order of
magnitude.
In the third column $\sigma_{\Delta}$ indicates the variance
of the night jumps.
These values are really very large and clearly show that there is a strong
discontinuity from the closing  price to the next day opening.
In the fourth column we show the variance of price fluctuation
within one day averaged over the 80 days
(average single day volatility).
Finally in the fifth column we show the variance of the price
fluctuations between two transactions.
One can see therefore that the night jumps are more than one order
of magnitude larger than the typical price change between two transactions.
This leads to a very serious problem if one tries to extend 
these time series beyond the time scale of a single day.
In fact, if one simply continues to the next day, one has anomalous jumps for
the night which cannot be treated as a standard price change.
An alternative possibility could be to artificially eliminate the night
jumps and rescale the price correspondingly.
This would produce a homogeneous data set which, however, does not correspond
to the original data.

This discussion clarifies that there is a fundamental
problem in extending the data beyond a single day.
Since the transactions within each day range from 500 to 3000,
this leads to  an important problem of finite size effects in relation
to the roughness exponent.
In the next section we are going to discuss these finite size effects and show
tat they are strongly amplified by the fat tail phenomenon.

\section{Roughness and Hurst exponent}
\begin{figure}[h]
\begin{center}
\resizebox{.75\columnwidth}{!}{\includegraphics{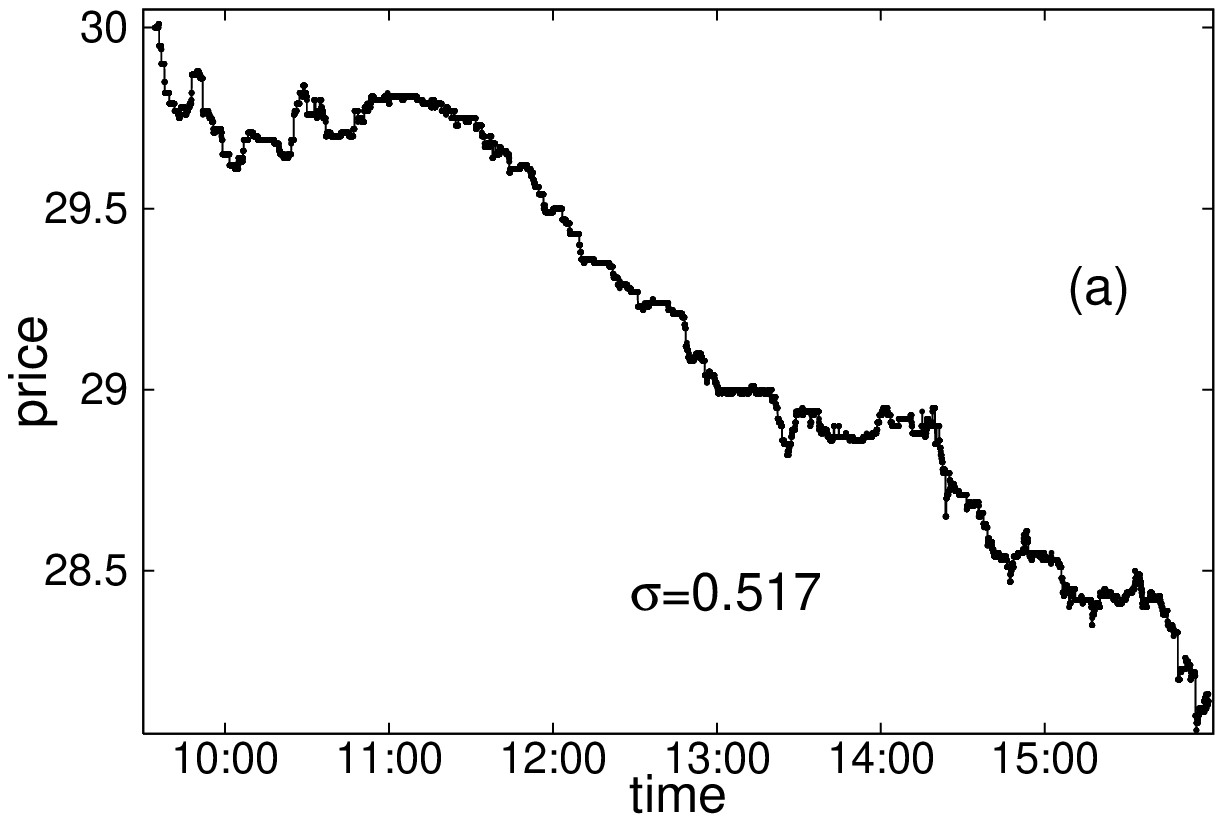}} \\
\resizebox{0.75\columnwidth}{!}{\includegraphics{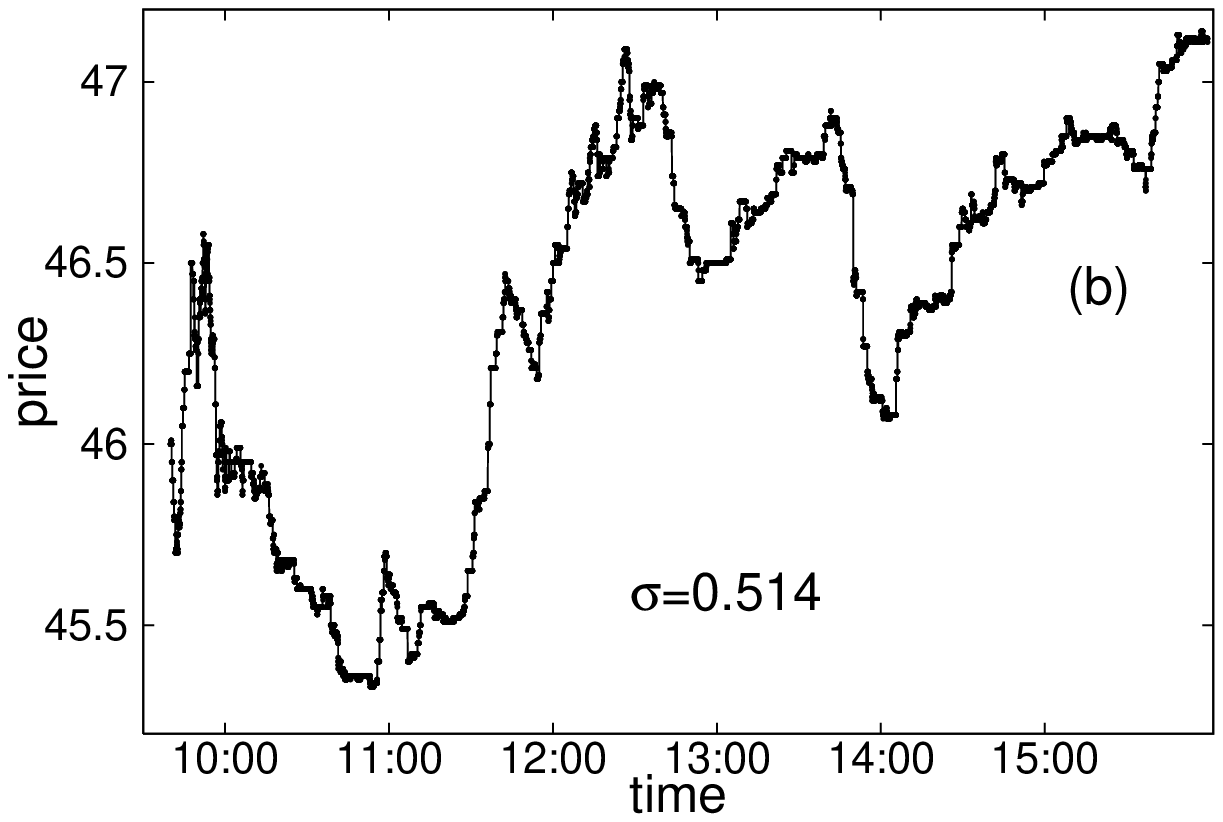}}
\caption{Examples of the day price dynamics of two stocks whose behavior, with
respect to the roughness, appears very different on a visual inspection. The
two stocks have a very similar variance ($\sigma$) for the price
distribution. Surprisingly, also the Hurst exponent will be similar for the
two cases.}
\label{fig3}
\end{center}
\end{figure}

The importance of a characterization of the roughness properties
is clearly illustrated in Fig. \ref{fig3}.
Here we see the behavior of the price of two stocks which are 
clearly very different with respect to their roughness properties.
The visual difference in roughness, however, does not influence the day
volatility $\sigma$, which is almost identical.
The idea is therefore to add new concepts to
characterize their different behavior.
We are going to see in the end that even the  Hurst exponent 
is not really optimal to this purpose and the challenge
of this new characterization should proceed along novel
lines which we outline at the end of the paper.

We first consider the problem of the characterization of the roughness in the
Hurst exponent including
the finite size effects.
The roughness exponent characterizes the scaling of the
price fluctuation   as a function of the size of the  interval
considered.

Originally this exponent was introduced for the time series of the levels
of the floods of the Nile river. The basic idea was to construct a profile
from these series and analyze its roughness.
This implied some peculiar construction which we can avoid because we have
the profile directly.

The characterization  of the roughness is complicated by the fact that
it corresponds to a problem of anisotropic scaling ~\citep{bib9}
and it can lead to  confusing results  in its practical applications~\citep{bib10}.
An example of these difficulties is illustrated by the fact that for the growth
of a rough profile the Renormalization Group procedure has to be implemented
in a rather sophisticated and unusual way~\citep{bib11}.
An illustration of this problem is also given by the fact that
the value of the fractal dimension of a rough surface is crucially 
dependent on the type of 
procedure one considers~\citep{bib11}.
The usual approach is to take the limit of small length scales for which
the relation between the dimension of  the profile, $D$, and the
Hurst exponent is~\citep{bib9}:
\begin{center}
\bea
D=2-H\;\;.
\label{eq1}
\eea
\end{center}
However, if one consider the limit of large scales
(not rigorous mathematically but often used in physics), 
one can get $D=1$ for the Brownian profiles which does not 
correspond any more to Eq. (\ref{eq1}).

In the data analysis one is forced to consider a finite interval and
necessarily
the two tendencies get mixed.
Even considering Eq. (\ref{eq1})
one can have various ambiguities.
In fact a large Hurst exponent corresponds to small value
of the fractal dimension which may appear strange.

Various problems contribute to this possible confusion.
The first is how one looks at a scaling law for an anisotropic problem.
The scaling for roughness links the vertical fluctuation $\Delta h$
as a function of the interval considered:

\bea
\Delta h (\Delta L) \sim {\Delta L}^H
\eea
In a physical perspective one has typically a lower cutoff and looks at the
behavior for large values of $\Delta L$.
Since for a random walk (Brownian profile) one has $H=1/2$ 
one could say that if $H>1/2$ this corresponds to a case which is more rough
than the Brownian profile.
However this is in apparent contradiction with Eq. (\ref{eq1})
because the value of $D$, if $H>1/2$, results smaller than the Brownian value
$(D=3/2)$.
This is because Eq. (\ref{eq1}) is derived in the limit $\Delta L \rightarrow
0$ in the spirit of the coverage approach to derive the fractal
dimension.

A similar confusion can be given by the existence of trends
in the dynamics of the system.
Consider for example a straight line behavior for which $\Delta h \sim \Delta
L$.
In this case one would have $H=1$ and $D=1$, namely the system is not rough
in the $\Delta L \rightarrow 0$ perspective but it is very rough in the 
$\Delta L \rightarrow \infty$ view.
In such a situation one should realize that a trend is present and that the
system is smooth. However, this distinction is not possible with the Hurst approach.

Actually in the real data one has an upper and a lower limit for $\Delta
L$, due to the intrinsic statistical limitation of the sample.
The exponent $H$ is then obtained by a fit in a certain range of  
scales and all the above problems are difficult to sort out.

\section{Roughness in a finite size Random Walk}

In this section we discuss the role of finite size effects in the determination
of the Hurst exponent. We start by deriving some analytical results for a
finite size random walk.
Consider the function:
\bea
\!\!\!\!{\mathcal R}(n)=\langle\max_{k=(ln+1),(ln+n)}(x_k)\, -
\min_{k=(ln+1),(ln+n)}(x_k)\rangle_l
\label{Neq1}
\eea
where $l=1,2,...,\frac Nn$ and 
$\{x_1,x_2,...,x_N\}$ are $N$ record in time of a variable $X$.
The function ${\mathcal R}(n)$ describes the expectation value of the
difference between maximum and minimum over an interval of size $n$.
${\mathcal R} (n)$, for many records in time
is very well described by the following empirical relation:
\bea
\mathcal R (n)\propto n^H
\eea
where $H$ is the Hurst exponent.
Now we want to check which is the effect of the finite
size in estimating the Hurst exponent.
To perform this analysis we consider a random walk and try to 
make an analytical calculation of the function $R(n)$.

Suppose that $\{\delta x_1, \delta x_2,...\}$ are independent random
variables, each taking the value 1 with probability $p$ and -1  otherwise.
Consider the sums :
\bea 
x_n=\sum_{i=1}^{n}\delta x_i
\eea
then the sequence $x=\{x_i:i \geq 0 \}$ is a simple random walk starting at the
origin.
Now we want to compute the expectation value of the maximum and the minimum
of the walk after $n$ steps.
In order to do that consider the Spitzer's identity which relates $\mathbb
E({M_n})$ to $\mathbb E \,({x_n^+})$ in the following way~\cite{bib12}:

\bea
\log \Big(\sum_{n=0}^{\infty}\,t^n \,\mathbb E(s^{M_n})\Big)=
\sum_{n=1}^{\infty}\frac 1n\, t^n \, \mathbb E \,(s^{x_n^+})        
\eea
where $M_n=\max\,\{x_i: 0\geq i \geq n \}$ is the maximum of the walk
up to time $n$, $x_n^{+}=\max\, \{0,x_n\}$, $s$ and $t$ are two 
auxiliary variables which absolute values are smaller than one and  $\mathbb E$ is the expectation value.
Considering the exponential of the Spitzer's identity and performing
the n-nth derivative we obtain:
\bea
\nn
\mathbb E\,(s^{M_n})&=&\frac 1n \sum_{k=1}^{n}\frac {\mathbb E
  (s^{x_n^+})}{(n-k)!}\,f^{(n-k)}{\big|}_0\;\;\;\;\; \mbox{where}\\
f(t)&=&\exp(g(t))\\
\nn 
g(t)&=&\sum_{n=1}^{\infty}\frac 1n\, t^n \, \mathbb E \,(s^{x_n^+})
\nn
\eea
and $f^{(n)}{\big|}_0$ is the $n$-nth derivative of $f(t)$ calculated in $t=0$.
For the derivative one can write a recursive expression:
\bea
f^{(n)}{\big|}_0=\sum_{k=1}^{n}\frac{(n-1)!}{(n-k)!}\mathbb E \,(s^{x_k^+})f^{(n-k)}{\big|}_0
\eea
The relation between $\mathbb E (M_n)$ and $\mathbb E (s^{M_n})$
for a symmetrical probability density function can be obtained by a 
straightforward calculation~\cite{bib12a}.
\bea
\nn
\mathbb E (s^{x_n^+})=1+\frac 12 \,\mathbb E (|x_n|)\,\ln(s)+{\mathcal O} {(\ln(s))}^2
\eea
By substituting this expression in that of $\mathbb E (s^{M_n})$
and  taking the limit for $s\rightarrow 1$:
\bea
\mathbb E (M_n)=\lim_{s\rightarrow 1}\frac{\mathbb E (s^{M_n})-1}{\ln(s)}
\eea
The basic final relation is therefore~\cite{bib12a}
\bea
\mathbb E (M_n)=\sum_{i=1}^{n}\frac{\mathbb E (|x_i|)}{2i}
\eea

\begin{figure}[h]
  \begin{center}
    \resizebox{0.75\columnwidth}{!}{\includegraphics{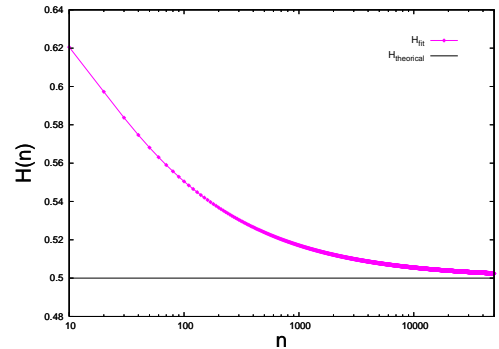}} 
    \caption{This plot shows the trend obtained fitting the curve 
${\mathcal  R}(n)$ for different values of the size $n$. The 
results shows a systematic overestimate of Hurst exponent for small size, due to finite
size effects. This is a general result and it shows that finite size effects
always enhance the apparent Hurst exponent. This enhancement can be understood
by considering that, in some sense, a single step would correspond to $H=1$, so
the asymptotic value $H=1/2$ is approached from above.}
    \label{Ngraf5}
  \end{center}
\end{figure}

From this expression it is possible to derive explicitly 
the expected value of the maximum as a function of the number of steps
of the walk. 
By considering that a similar expression holds also for the minimum,
one can directly compute the effective Hurst exponent for random walk
of any size. The specific example of Gaussian increments or $\pm 1$
increments are considered in detail in Ref.~\cite{bib12a}.
Replacing the results obtained for $\mathbb E (M_n)$ in the expression
of ${\mathcal  R}(n)$ we can plot the average span as a function of $n$
and execute a fit to estimate the value of $H$.
Executing a fit in the region $[10,1000]$ we obtain a value of the slope
that is grater then the asymptotic one.
In Fig. \ref{Ngraf5} we shows the result for the effective Hurst exponent that we have obtained performing
the fit in the region $[10,n]$ for the random walk with two identical steps
$(\pm 1)$. One can see that finite size effects are very important 
and a seriously affect the apparent value of $H$.

\begin{figure}[h]
  \begin{center}

    \resizebox{0.75\columnwidth}{!}{\includegraphics{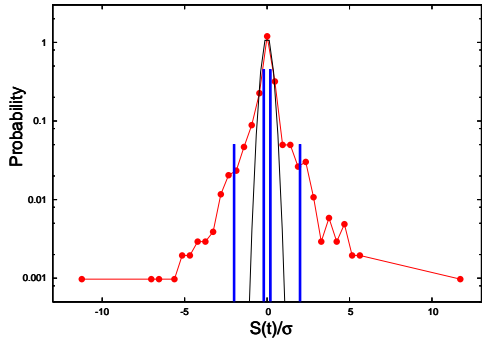}} 
    \caption{Probability function for high frequency
price difference of the BRO stock during a day. The solid
line is the Gaussian fit of the data.
The boxes represent a model to estimate the effect of the 
tails for the random walk. The probability is estimate
by an histogram given by a value $\pm \epsilon$ which 
has $0.45$ of probability an a tail $\pm 10\epsilon$ 
with probability $0.05$.
In this plot $\epsilon=0.2$.}
    \label{Ngraf6}
  \end{center}
\end{figure}

The random walk models considered until now have a distribution of individual
steps corresponding to a Gaussian distribution or to two identical steps. Real
price differences however, are characterized by a distribution of sizes which
strongly deviates from these (``fat tails``). 
For example, if we consider the histogram of the quantity:
\bea
S(t)=\ln P (t+1) - \ln P(t) ,
\eea
we find a distributions with large tails, as shown  in Fig. \ref{Ngraf6}.

\begin{figure}[h]
  \begin{center}
    \resizebox{0.75\columnwidth}{!}{\includegraphics{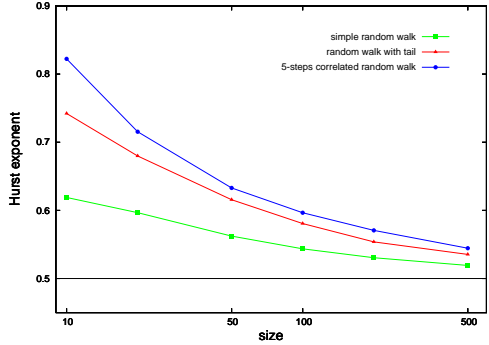}} 
    \caption{The value of $H(n)$ is shown for three
different random walk models: normal, correlated
and with fat tails. Finite size effect already
present in the normal random walk are amplified
by the presence of fat tails and correlations.
In the x-axis is plotted the effective size, that is $\frac n{10}$. The values
are averaged over $1000$ realizations.}
    \label{Ngraf7}
  \end{center}
\end{figure}

To analyze the effect of the fat tails in the 
evaluation of the Hurst exponent, we can consider a
model of random walk with increments that take the values
$\delta x =\pm \epsilon$ with probability $0.45$ and $\delta x=\pm 10\epsilon$
with probability $0.05$.
The histogram in Fig. \ref{Ngraf6} represent such a model.
We have performed a numerical analysis
of the Hurst exponent for a random walk with fat tails
to study their role on the finite size effects.
To this purpose we have generated $1000$  random walks  of this kind of size $n$ 
with $n=[100:5000]$ and we have calculated the function $R(n)$
for each sample.
After calculating the average of $R(n)$, we have considered
the plot $R(n)$ as a function of $n$ and 
the evaluation of $H(n)$ has been performed in the region
$[\frac n{100};\frac n{10}]$.
Figure \ref{Ngraf7} shows the result obtained, a comparison
with a normal and a correlated random walk is also shown.

The fact that fat tails and correlations enhance the finite size effects is
easy to understand. In case of correlated random walks the effective 
number of independent steps is strongly reduced. In the case of fat tails
instead only the tails give the main contribution to the profile.

This findings could also have implications for very long times if combined
with the non stationarity of the price dynamics.
It should be  considered the possibility that even the asymptotic regime is
still altered by these effects.
This could suggest a different interpretation of the deviation of $H$  from
the value $1/2$, which is usually proposed in terms of long range correlations~\cite{bib8}.

The Fig. \ref{Ngraf7} shows the inefficiency of the 
Hurst exponent's approach to the study of the roughness
for systems with a small size.
The results are clearly affected by the effect of a finite size
and the interpretation of $H>1/2$ as a long range correlation
could be misleading.

\section{Analysis of NYSE stocks}

\begin{figure}
  \begin{center}
       \resizebox{0.75\columnwidth}{!}{\includegraphics{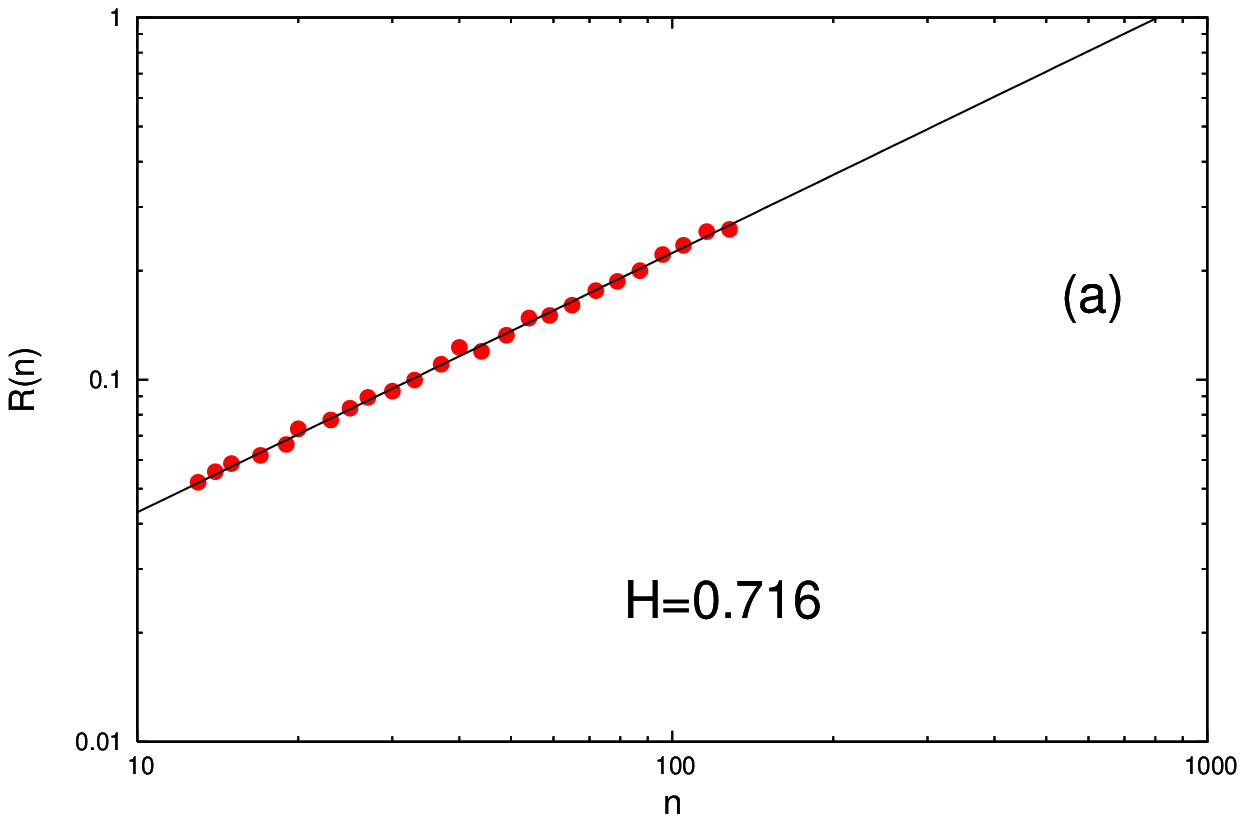}} \\
      \resizebox{0.75\columnwidth}{!}{\includegraphics{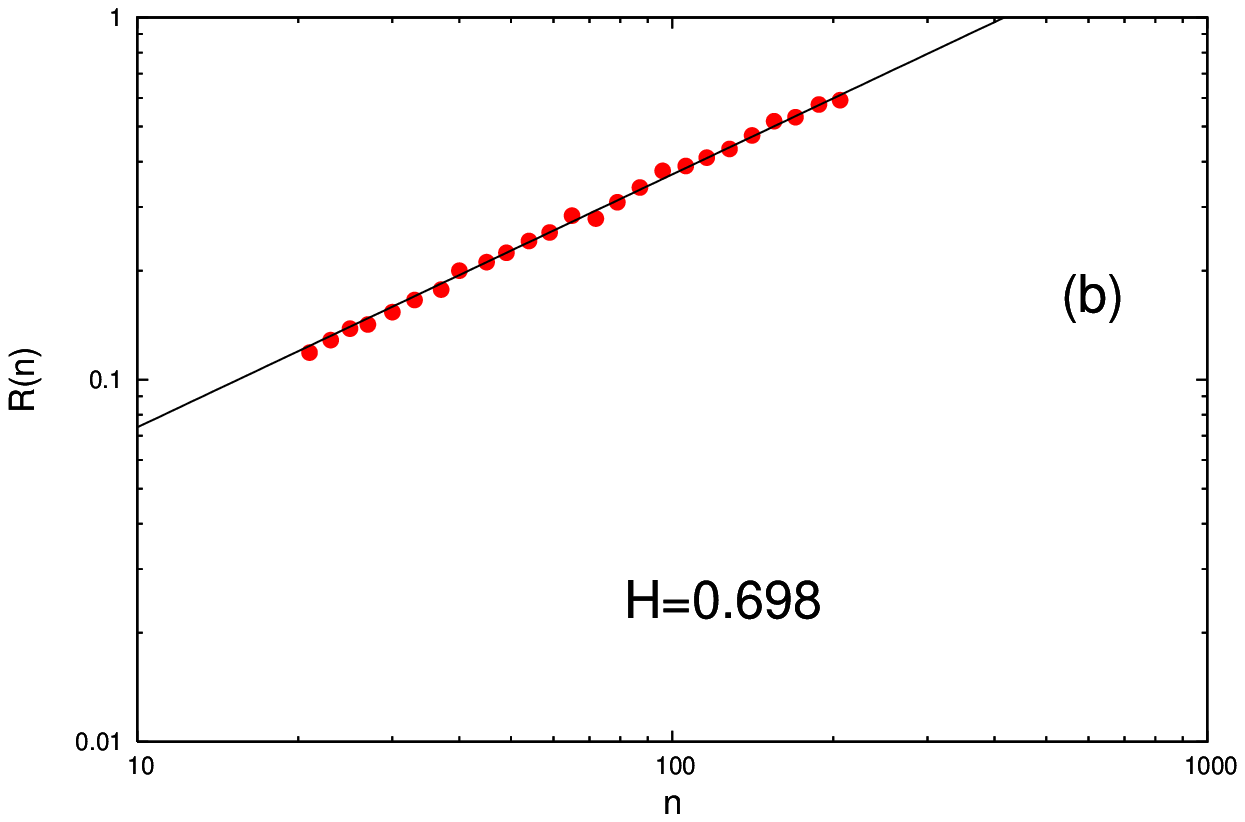}} 
     \caption{Analysis in terms of the Hurst exponent of the two stocks shown
      in Fig. \ref{fig3}. The case (a) refers to the stock which appears
      smooth, while (b) is the other one. One can see that the value of $H$
is very similar despite the apparent differences between the two behavior
      (Fig. \ref{fig3}).}
    \label{2Hurst}
  \end{center}
\end{figure}

First we consider the Hurst analysis for the two stocks plotted in
Fig. \ref{fig3} and the relative results are shown in Fig. \ref{2Hurst}.

The values of the two exponents $H$ are very similar in spite of
the large difference of the two stocks in their apparent roughness properties.
This shows that the exponent $H$ is not suitable to characterize
the different roughness properties of the two stocks.
\begin{table}[h]
\caption{Hurst exponent for 20 NYSE stocks. $H$ is the average daily value
  over the 80 days. $H_{max}$ and $H_{min}$ are the maximum and the minimum
and $\sigma$ the variance. $<N>$ is the average number of transactions per day.} 
\label{tabhurst}
\begin{tabular}{lccccc}
\hline\noalign{\smallskip}
stock & $H$ &  $H_{max}$  & $H_{min}$ & $\sigma$ & $<N>$ \\ 
\noalign{\smallskip}\hline\noalign{\smallskip}

AH &  0.599 & 0.732 & 0.489  & 0.0215 & 1535.77\\ 

AVO &  0.615 & 0.785 & 0.501  & 0.0170 & 1296.71\\ 

BA &  0.573 & 0.694 & 0.478  & 0.0106 & 3323.37\\ 

BRO &  0.662 & 0.792 & 0.557  & 0.0161 & 853.91\\ 

CAI &  0.641 & 0.751 & 0.478  & 0.0232 & 1052.58\\ 

DRI &  0.575 & 0.699 & 0.445  & 0.0106 & 1446.65\\ 

GE &  0.526 & 0.653 & 0.406  & 0.0065 & 5598.83\\ 

GLK &  0.627 & 0.780 & 0.484  & 0.0105 & 1114.01\\ 

GM &  0.574 & 0.677 & 0.462  & 0.0083 & 3405.84\\ 

JWN &  0.579 & 0.738 & 0.457  & 0.0125 & 2025.67\\ 

KSS &  0.570 & 0.686 & 0.438  & 0.0135 & 2789.09\\ 

MCD &  0.559 & 0.691 & 0.417  & 0.0076 & 3480.63\\ 

MHS &  0.612 & 0.750 & 0.460  & 0.0113 & 1792.51\\ 

MIK &  0.591 & 0.752 & 0.456  & 0.0132 & 1377.84\\ 

MLS &  0.635 & 0.914 & 0.496  & 0.0204 & 759.27\\ 

PG &  0.551 & 0.662 & 0.456  & 0.0091 & 4135.80\\ 

TXI &  0.636 & 0.776 & 0.473  & 0.0296 & 733.68\\ 

UDI &  0.679 & 0.781 & 0.524  & 0.0147 & 774.25\\ 

VNO &  0.622 & 0.777 & 0.506  & 0.0244 & 883.78\\ 
\noalign{\smallskip}\hline
\end{tabular}
\end{table}

\begin{figure}[h]
  \begin{center}
    \resizebox{0.75\columnwidth}{!}{\includegraphics{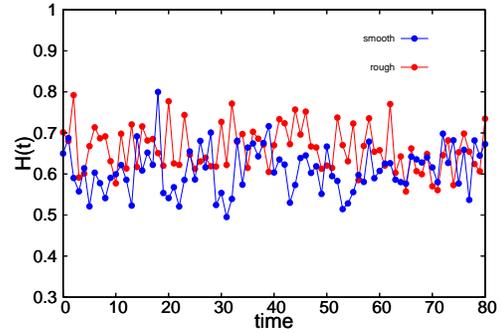}} 
    \caption{Time dependence of the Hurst exponent $H(t)$ for the two stocks
      shown in Fig. \ref{fig3}.}
    \label{Htime}
  \end{center}
\end{figure}

We then consider the entire series of $20$ stocks and the results are
reported in Tab. \ref{tabhurst}. Here $H$ represents the a daily  value
averaged over 80 days.
Then $H_{max}$ and $H_{min}$ are the maximum and the minimum values
respectively, $\sigma$  is the variance averaged over the 80 values and $<N>$
is the average number of transactions per day.
In Fig. \ref{Htime} we report the time behavior of $H(t)$ for the 80 days for
the two stocks of Fig. \ref{fig3}.
With respect to previous analysis of the time dependence of H(t)~\cite{bib7},
we can observe that the daily variability of single stocks is much 
larger than that of global indices over long times.
In addition also the average is appreciably larger.

A  general result is that the value of $H$ is systematically larger than
$1/2$. The usual interpretation would be to conclude that long range
correlations are present~\cite{bib8}.
However, in view of our previous discussion we would instead propose that this
deviation from $1/2$ is precisely due to finite size effects, combined with
the fat tail phenomenon.
A further support to this interpretation is that if we built a long time
series by eliminating the night jumps, one observes a convergency towards the
value $1/2$.
Also one may note that stocks with a relatively large number of transactions 
per day ($<N>$), like for example GE stock, are much closer to the random walk
value $H=1/2$.

The fact that apparently different profiles with respect to the roughness
lead to value of $H$ which are very similar is due to a variety of reasons.
The overall enhancement with respect to the standard value $1/2$ is, in our
opinion,
mostly due to the finite size effects phenomenon.
However, this does not explain why two profiles which appear very different,
like those in Fig. \ref{fig3}, finally, lead to very similar values of $H$.
This is probably due to the fact that the Hurst approach tends to mix
the role of trends with fluctuations and in the next section we are going to
propose a different method to resolve this problem.

To complete our analysis, we consider the generalized Hurst exponent
in the spirit of Ref.~\cite{bib14}.
To this purpose we analyze a q-th order price difference  correlation
function defined by:

\bea
G_q(\tau)=<|p(x)-p(x+\tau)|^q>^{\frac 1q}
\eea

The generalized Hurst exponent $H_q$ can be defined from
the scaling behavior
of $G_q(\tau)$:

\bea
G_q(\tau)\sim {\tau}^{H_q}
\eea

For a simple random walk $H_q=H=1/2$ independently of $q$.
We have calculate the function $G_q(\tau)$ for the two test-stocks.

\begin{figure}
  \begin{center}
      \resizebox{0.75\columnwidth}{!}{\includegraphics{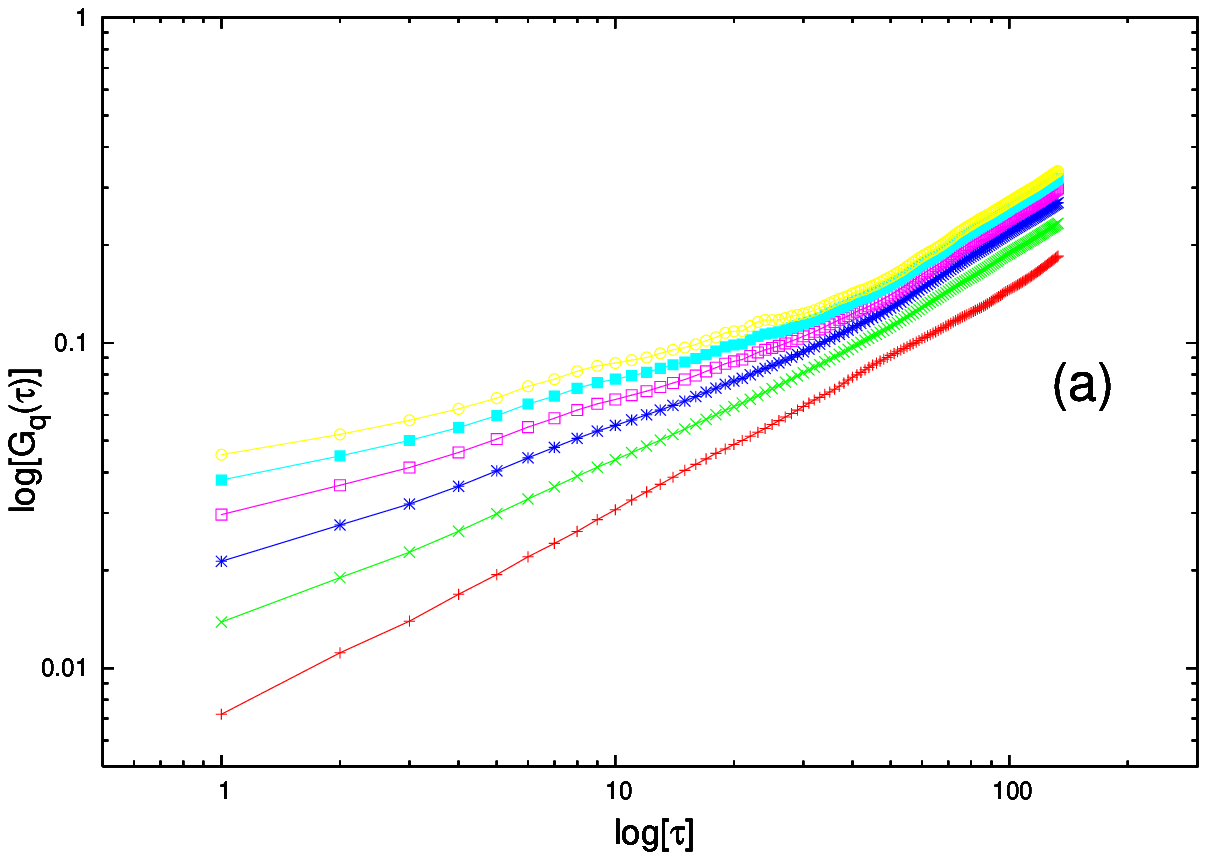}} \\
      \resizebox{0.75\columnwidth}{!}{\includegraphics{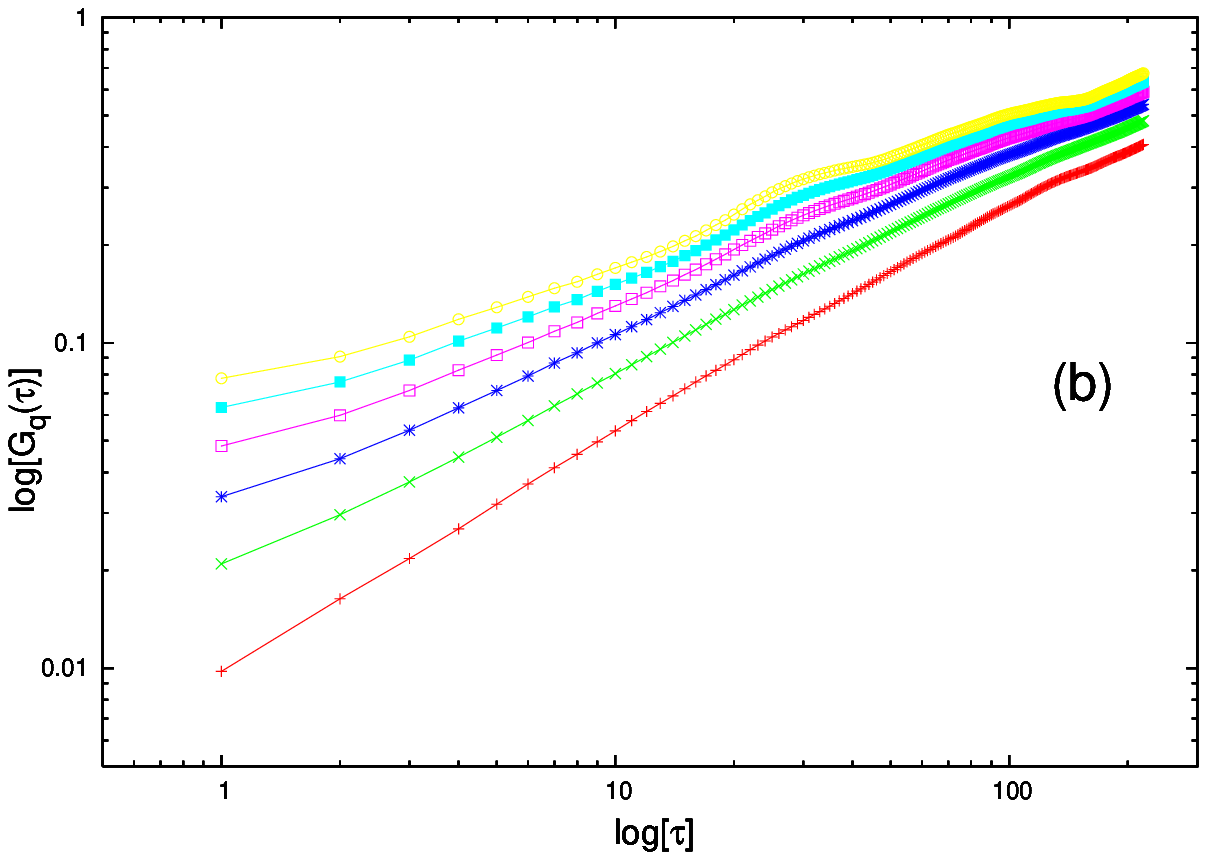}} 
    \caption{$G_q(\tau)$ as a function of $\tau$ in a log-log scale
for the two test-stock ((a) is the smooth and (b) the rough one). In both (a)
and (b), from bottom to top $q=1,2,3,4,5,6$.}
\label{grafq}
\end{center}
\end{figure}

The results are shown in Fig. \ref{grafq}
and  show that $H_q$  is not a constant  but strongly depends on $q$.
This result provides an evidence that the characteristics of the profile are
dominated by the large jumps due to the fat tail properties.

\section{New approach to roughness as fluctuation from Moving Average}

In this section we consider a new method to characterize the roughness.
The basic idea is to be able to perform an automatic detrendization of the price signal.
This can be achieved by the difference between the price variable
and its moving average defined in an optimal way.
At each transaction point $t_i$ we define the moving average of the price
$P(t_i)$, with a characteristic time $\tau$, as:
\bea
P_\tau(t_i)=\frac 1{N_\tau}\sum_jP(t_j)
\eea
where $N_\tau$ are the number of transactions in the time interval
$[-\tau/2:\tau/2]$.
This function corresponds to the symmetric average over an interval
of size $N_\tau$ around $t_i$.

\begin{figure}[h]
  \begin{center}
    \resizebox{0.75\columnwidth}{!}{\includegraphics{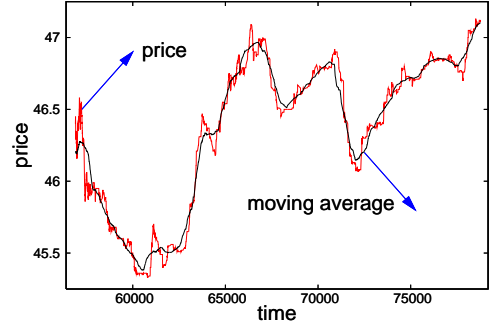}} 
    \caption{Example of price fluctuations and the corresponding moving
average. In our case we consider a symmetrized moving average defined
as the average of the price over a symmetric interval of total size
$\tau$. }
    \label{mmobile}
  \end{center}
\end{figure}

One can then consider the maximum deviation of $P(T_i)$ from $P_\tau(t_i)$
over an interval of a certain size, in our case we consider a single day:
\bea
R_\tau=\max_i |P(t_i)-P_\tau(t_i)|
\label{eq15}
\eea
This may appear similar to the standard definition of roughness
which gives the absolute fluctuation in a time interval $\tau$.
Instead the use of $R_{\tau}$ corresponds to an automatic
  detrendization which appears more appropriate to study the
roughness.
Our approach is similar to the one of Ref.~\cite{bib13}, 
but with the difference that we use a symmetrized definition
of the moving average while Ref.~\cite{bib13}
defines the moving average only with respect to a previous
time interval.

\begin{figure}
  \begin{center}
       \resizebox{0.75\columnwidth}{!}{\includegraphics{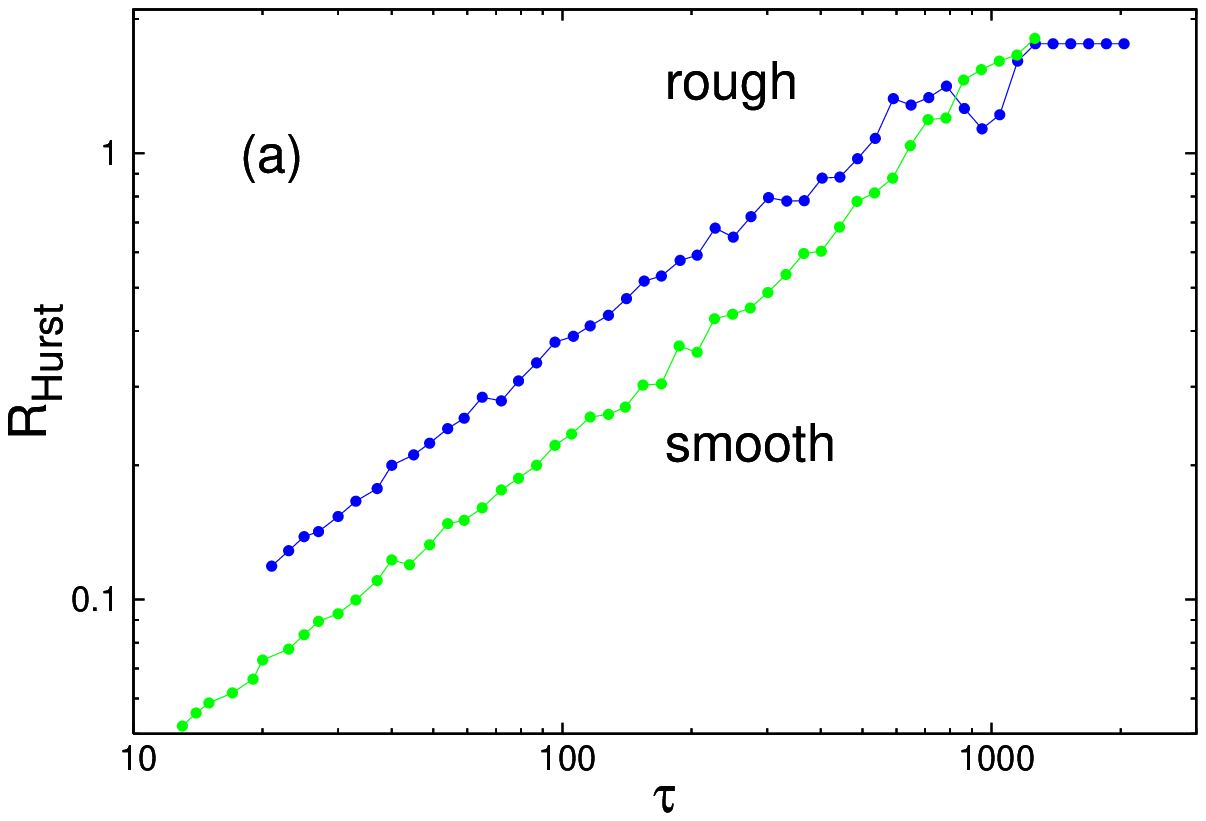}} \\
      \resizebox{0.75\columnwidth}{!}{\includegraphics{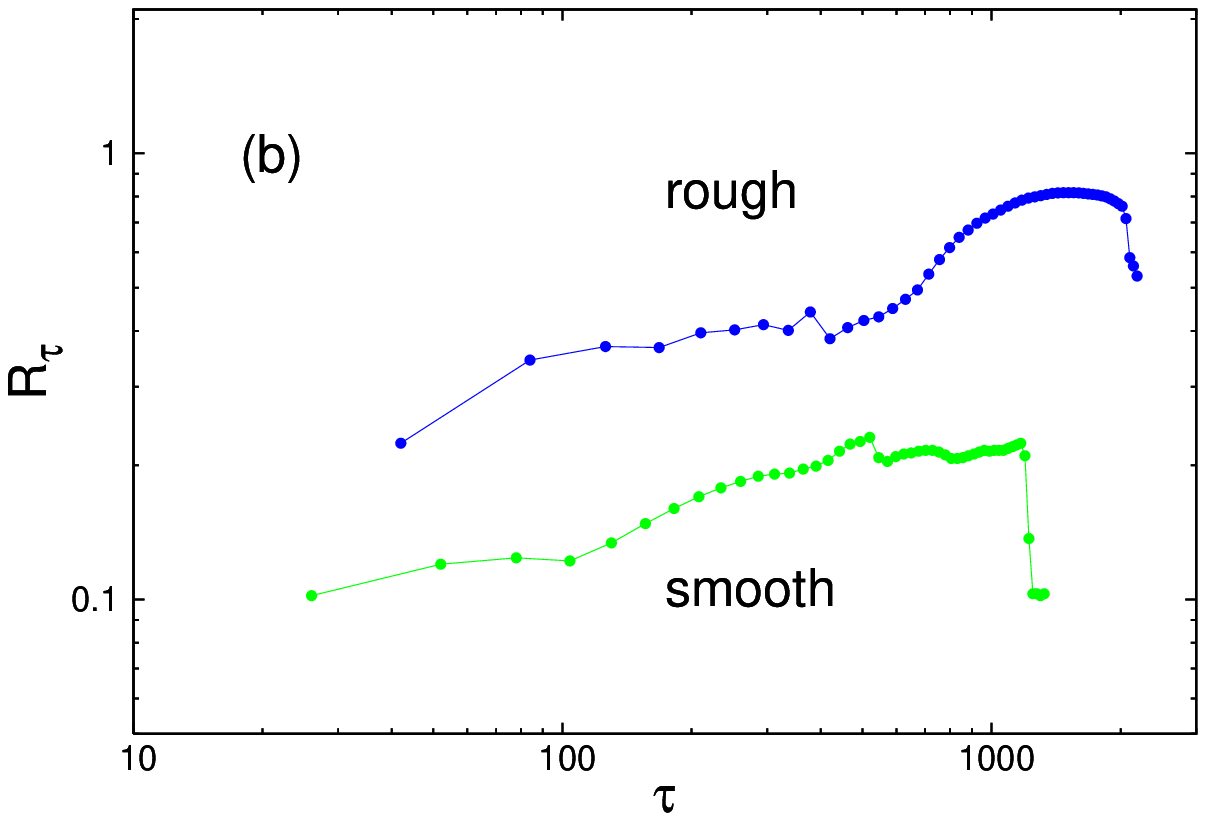}} 
    \caption{(a) Fluctuations over intervals of different size ($\tau$)
defined by the differences between maximum and minimum values over intervals
of size $\tau$. These curves were used in the previous sections to compute
the Hurst exponent in the standard way. The two curves refer to the
two stock of Fig. \ref{fig3}. One can observe that the slope 
is similar for the two cases and, at relatively large scale ($\tau$) even
the amplitudes become rather similar.
(b) In this case the amplitudes are defined by the fluctuations from the
moving averages as given by Eq. (\ref{eq15}). In this case there is 
a marked difference in slope and even more in amplitude. This example
clarifies that this new definition of roughness can be more
useful to classify the stock dynamics.
}
\label{2ampi}
\end{center}
\end{figure}

In Fig. \ref{2ampi} we show the values of $R_\tau$ for the two stocks shown in
Fig. \ref{fig3} and, for comparison,  the same stocks analyzed with the Hurst
method. One can see that the fluctuations from the moving average are more 
appropriate to describe the difference between these stocks which cannot be
detected with the standard Hurst approach.

\section{Discussion and Conclusions}

We have considered the roughness properties as a new element to 
characterize the high frequency stock-price fluctuations.
The data considered include all transactions and show a large
night jump between one day and the next. For this reasons the dataset are
statistically homogeneous only within each day. This leads to a 
serious problem of finite size effects which we have  analyzed by using various 
random walk models as examples. We have computed the effective Hurst exponent as a 
function of the size of the system. The basic result is that the 
finite size effects lead to a systematic enhancement of the
effective Hurst exponent and this tendency is amplified
by the inclusion of fat tails and eventual correlations.

An analysis of real stock-price behavior leads to the conclusion that
most of the deviations from the random walk value ($H=1/2$) are
indeed due to finite size effects. Considering the 
importance of non-stationarity phenomenon one may conjecture that the
finite size effects could be important even for long series of data.

Concerning the roughness analysis we conclude that the
standard Hurst approach is not very sensitive in order to characterize the 
various stock-price behaviors. We propose a different roughness analysis based
on the fluctuations from a symmetrized moving average. This has the advantage
of an automatic detrendization of the signal without any {\itshape {ad hoc}}
modification of the original data. This new method appears much more
useful than the standard one in order to characterize the 
fluctuations behavior of different stock as shown clearly by the analysis
of the two cases in Fig. \ref{fig3}.


\begin{thebibliography}{99}

\bibitem{bib1}
B.~Mandelbrot, \textit{Fractals and Scaling in Finance} (Springer Verlag, New York, 1997).
\bibitem{bib2}
R.N.~Mantegna, H.E.~Stanley, \textit{An Introduction to Econophysics} (CambridgeUniversity Press, Cambridge, 2000). 

\bibitem{bib3}
J.P.~Bouchaud, \textit{Theory of Financial Risk}, (Cambridge
University Press, Cambridge, 2000).

\bibitem{bib4}
H.E.~Hurst, 
Transaction of the  American Society of Civil Engineers \textbf{116},  770-808
(1951).

\bibitem{bib5}
S.O.~Cajueiro, B.~Tabak, 
 Physica A \textbf{336}, 521-537 (2004).

\bibitem{bib6}
D.~Grech, Z.~Mazur, 
 Physica A  \textbf{336}, 133-145 (2004).

\bibitem{bib7}
A.~Carbone, G.~Castelli, H.E.~Stanley, 
Physica A \textbf{344}, 267-271 (2004).



\bibitem{bib8}
T.~Di Matteo, T.~Aste, M.M.~Dacorogna, 
Journal of Banking and Finance \textbf{29}, 827-851 (2005).


\bibitem{bib9}
A.L.~Barabasi, H.E.~Stanley, \textit{Fractal Concepts in Surface Growth} (Cambridge
University Press, Cambridge, 1994).

\bibitem{bib10}
L.~Pietronero, \textit{Order and Chaos in Nonlinear Physical System}, edited by S. Lundqvist, N. H. March, and M. Tosi (Plenum Publishing Corporation, New York, 1988), 227.

\bibitem{bib11}
C.~Castellano, M.~Marsili, L.~Pietronero, Phys. Rev. Lett. \textbf{80}, 3527 (1998).

\bibitem{bib12}
G.~Grimmet, D.~Stirzaker, \textit{Probability and Random Processes}, (Oxford University
Press, Oxford, 2001).

\bibitem{bib12a}
V.~Alfi, F.~Coccetti, M.~Marotta, A.~Petri, L.~Pietronero, 
Physica A, in print 2006, cond-mat/0601230.

\bibitem{bib14}
J.~Asikainen, S.~Majaniemi, M.~Dubé, T.~Ala-Nissila, 
Phys. Rev E  \textbf{65}, 052104 (2002).


\bibitem{bib13}
E.~Alessio, A.~Carbone, G.~Castelli, V.~Frappietro, 
Eur. Phys. J. B \textbf{27}, 197-200 (2002).


\end{thebibliography}
\end{document}